\documentclass[preprintnumbers,showpacs,amsmath,amssymb,twocolumn,pra]
{revtex4-1}
\usepackage{color}
\usepackage{amssymb}
\usepackage{amsmath}
\usepackage{epsfig}
\usepackage{ulem}
\usepackage{mathrsfs}

\newcommand{\ehbar}{\hbar_{\text{eff}}}

\newcommand{\mtn}{\mathcal{N}}

\begin{document}

\title{Directed momentum current induced by the $\mathcal{PT}$-symmetric driving}

\author{Wen-Lei Zhao$^{1}$}
\email[]{zwlphys@aliyun.com}
\author{Jiaozi Wang$^{2}$}
\email[]{wangjz@ustc.edu.cn}
\author{Xiaohui Wang$^{3}$}
\author{Peiqing Tong$^{4,5}$}
\email[]{pqtong@njnu.edu.cn}

\affiliation{
$^1$School of Science, Jiangxi University of Science and Technology, Ganzhou 341000, China\\
$^2$Department of Modern Physics, University of Science and Technology of China, Hefei 230026, China\\
$^3$School of Information Engineering, Jiangxi University of Science and Technology, Ganzhou 341000, China\\
$^4$Department of Physics and Institute of Theoretical Physics, Nanjing Normal University, Nanjing 210023, China\\
$^5$Jiangsu Provincial Key Laboratory for Numerical Simulation of Large Scale Complex Systems, Nanjing Normal University, Nanjing, Jiangsu 210023, China
}

\begin{abstract}
We investigate the directed momentum current in the quantum kicked rotor model with $\mathcal{PT}$ symmetric deriving potential. For the quantum non-resonance case, the values of quasi-energy become to be complex when the strength of imaginary part of the kicking potential exceeds a threshold value, which demonstrates the appearance of the  spontaneous $\mathcal{PT}$ symmetry breaking.
In the vicinity of the transition point, the momentum current exhibits a staircase growth with time.
Each platform of the momentum current corresponds to the mean momentum of some eigenstates of the Floquet operator whose imaginary parts of the quasi-energy are significantly large.
Above the transition point, the momentum current  increases linearly with time. Interestingly, its acceleration rate exhibits a kind of ``quantized'' increment with the kicking strength. We propose a modified classical acceleration mode of the kicked rotor model to explain such an intriguing phenomenon. Our theoretical prediction is in good agreement  with numerical results.
\end{abstract}


%
%
%

\maketitle


\section{Introduction}

The directed transport of macroscopic particles has attracted intensive attentions in the past few decades~\cite{Hanggi09,Gong06,Pelc09,Gong04,WangJiao08}. The phenomenon implies the flow of energy or the transmission of information. Therefore, its mechanism is significantly important for the design of quantum heat engines and quantum information. It has also wide applications in  the construction of nanoscale devices, such as particle separation and electron pumps, and for the understanding of biological molecular
motors~\cite{Kohler05,Julicher97,Lehmann02}. Formation of directed current is closely related to the breaking of spatiotemporal symmetry or the topological property of system. These two kinds of features are controllable in the Floquet-driven system by manipulating the external potential.
For example, the spatially-nonsymmetric driving potential leads to the directed motion of cold atoms in optical lattice~\cite{Sadgrove07,Dadras18,Summy16,Ni17}. A double-kicking extension of quantum kicked rotor (QKR) model exhibits topological momentum current~\cite{Ho2012}.
More recently, the atom-optics experiments of QKR model has reported the directed acceleration in momentum space~\cite{Sadgrove13}.
Indeed, the system of cold atoms driven by time-periodical optical lattice is
an ideal platform for investigating the directed transport
phenomenon~\cite{Kenfack08,Lundh05,Zhao14}.

On the other hand, the quantum dynamics with $\mathcal{PT}$ symmetry is fundamentally important~\cite{Konotop16,Bender11,Mostafazadeh10,Kozlov15}. This field has been regarded as a significant extension of traditional Hermitian systems~\cite{Bender98}. A unique property of $\mathcal{PT}$ symmetric quantum systems is a spontaneous  transition that is the real energy eigenvalues
become to be complex when the strength of the imaginary part of the complex potential exceeds a threshold value.
Interestingly, the experimental progress has realized the ultracold atoms in complex
optics lattices~\cite{Keller97,Xiao16,Li16,Kreibich16,Kreibich13,Kreibich14,Hang18}, which opens the opportunity for investigating the transport behavior in the $\mathcal{PT}$ symmetric potential.
Previous studies on wavepackets dynamics of these system show that, with the $\mathcal{PT}$ symmetry being broken, the energy band
merging leads to peculiar transport behavior of optic wave
and matter wave, such as double refraction,  nonreciprocal
diffraction,  bifurcation, and many others~\cite{Makris08,Longshi09}.

The extension of Floquet-driven system to $\mathcal{PT}$
symmetric regime induces exotic physics~\cite{West2010}. More recent study on a QKR model with $\mathcal{PT}$ symmetry shows that the spontaneous
$\mathcal{PT}$ transition occurs in the dynamical localization regime, while the $\mathcal{PT}$ symmetry is always broken in the quantum resonance case~\cite{Longshi2017}. Moreover, they discover that the momentum current is unboundedly accelerated in quantum resonance, and that it is suppressed by dynamical localization in condition that the $\mathcal{PT}$ symmetry is preserved. It is known that the Talbot effect of quantum resonance case causes the periodical revival of the QKR wavepackets, therefore the system experiences a constant force which facilitates the directed current. On the other hand, the transport behavior of matter wave in quantum non-resonance situation is still an open topic, which needs urgent investigation.

Motivated by these studies, we investigate the directed acceleration of wavepackets in the quantum non-resonance case via a QKR model with $\mathcal{PT}$ symmetry.
We find that the breaking of the $\mathcal{PT}$ symmetry can induce rich transport behaviour in momentum space. Specifically, in the vicinity of transition points, the mean momentum exhibits the staircase growth with time.
Detailed investigation reveals that the platform of the momentum current is  determined by mean momentum of some eigenstates of the Floquet operator with large imaginary parts of quasi-energy.
Moreover, it is found that the momentum of those eigenstates concentrate on several separate values which are in one-to-one correspondence to the momentum of each platform.
Above the transition points, the mean momentum  increases linearly with time. More interesting is that the acceleration rate shows a ``quantized'' increment with the kick strength. The underlying physics behind such an interesting phenomenon is the modification of the classical acceleration mode of kicked rotor model by the gain-or-loss mechanism of the complex kicking potential. Our theoretical prediction of the acceleration rate of such ``quantized'' momentum current is in perfect consistence with numerical results.

Conventionally, it is believed that, in the quantum non-resonance case, the mechanism of dynamical localization suppresses directed motion of microscopic particles. Our finding of the directed momentum current in the dynamical localization regime which induced by breaking of $\mathcal{PT}$ symmetry may sight a new light on this fundamental problem.
The experimental advances in atom-optics had made
it possible to realize the complex optical lattice.
We therefore hope our theoretical results will stimulate future experiments in this field.

The paper is organized as follows. In Sec.~\ref{sec2}, we describe the system and show the directed momentum current.
In Sec.~\ref{sec3}, we study the  ``quantized'' acceleration mode. Summary is presented in Sec.~\ref{sec4}.

\section{Directed momentum current}\label{sec2}

We consider the QKR model with the $\mathcal{PT}$-symmetry for which the Schr\"odinger equation
takes the form
\begin{equation}\label{schorEq0}
i\hbar \frac{\partial \psi}{\partial t}= \left\{ \frac{\hat{P}^2}{2 I} + V_0\left[\cos(\theta) + i \lambda \sin(\theta)\right]\delta_T  \right \} \psi\;,
\end{equation}
where $\hat{P}$ is the angular momentum operator, $\theta$ is the angle coordinate, $I$ is the moment of inertia, $V_0$ is strength of
the kicking potential with $\lambda$ being the strength of its imaginary component,
and $\delta_T=\sum_n
\delta(t-nT) $ with $T$ being the kick period.
The time evolution of quantum states during one period is governed by the Floquet operator
\begin{equation}\label{evol}
U =  \exp\left[-i\frac{V_0(\theta)}{\hbar}\right]\exp\left( \frac{-i}{\hbar}\frac{\hat{P}^2T}{2 I}\right)\;,
\end{equation}
where $V_0(\theta) = V_0\left[ \cos(\theta) + i \lambda \sin(\theta)\right]$~\cite{Longshi2017}.

In the angular momentum representation, i.e., $\hat{P} |\varphi_n \rangle = n\hbar | \varphi_n \rangle$ with $\langle \theta \vert \varphi_n \rangle = e^{i n\theta}/\sqrt{2\pi}$, the
free evolution operator, namely, the second term in the right hand side of Eq.~\eqref{evol}, can be written as
\begin{equation}\label{FEvol}
U_f =  \exp\left( -i\frac{n^2 \hbar T}{2 I}\right)\;.
\end{equation}
It is evident that the $U_f$ is determined by a dimensionless factor $\hbar T/I$. Hence, for the convenience of investigation, we define it as an effective planck constant $\ehbar = \hbar T/I$, for which the dimensionless angular momentum is $p_n = n \ehbar$. Accordingly, we have made the scaling for the angular momentum operator as $\hat{p}=\hat{P}T/I$, for which the eigenequation is $\hat{p} |\varphi_n \rangle = p_n | \varphi_n \rangle$.
Then, the free evolution operator can be expressed as \begin{equation}\label{FEvol2}
U_f =  \exp\left( -\frac{i}{\ehbar}\frac{\hat{p}^2}{2 }\right)\;.
\end{equation}

With the effective Planck constant, the kicking evolution operator, namely, the first term in the right hand side of Eq.~\eqref{evol}, is rewritten as
\begin{align}\label{KEvol}
U_K &=\exp\left[-i\frac{V_0(\theta)}{\hbar}\right]=\exp\left[-i\frac{TV_0(\theta)}{\ehbar I}\right]\;,\\\nonumber
&=\exp\left[-i\frac{V_K(\theta)}{\ehbar }\right]\;,
\end{align}
where $V_K (\theta) = K\left[ \cos(\theta) + i \lambda \sin(\theta)\right]$ with the dimensionless  kicking strength $K = V_0 T /I$. The reason for introducing this dimensionless kick strength $K$ is that it is the only parameter controlling the classical dynamics governed by the well-known mapping equation~\cite{Izrailev87}. Traditional investigations on the quantum-classical correspondence of such a chaotic system mainly concentrates on the case that the $K$ is a constant, for which the classical limit is fixed. With the decrease of  $\ehbar$, i.e.,  $K/\ehbar \rightarrow \infty$, the quantum dynamics will be consistent with its classical counterpart for long enough time~\cite{Haake04}. In the present work, we indeed find that the spreading of wavepakcets of the non-Hermitian kicked rotor follows the classical acceleration modes in condition that $K/\ehbar \gg 1$.

By combining Eqs.~\eqref{FEvol2} and ~\eqref{KEvol}, the Floquet operator in dimensionless units reads
\begin{equation}\label{evol2}
U =  \exp\left[-i\frac{V_K(\theta)}{\ehbar }\right] \exp\left( -\frac{i}{\ehbar}\frac{\hat{p}^2}{2 }\right)\;.
\end{equation}
The eigenequation of the Floquet operator reads
\begin{equation}
U|\psi_{\varepsilon}\rangle =  e^{-i \varepsilon}|\psi_{\varepsilon}\rangle\;,
\end{equation}\label{EEQFloqoper}
where $\varepsilon$ indicates the quasi-energy.
Quantum nonresonance corresponds to irrational values of $\ehbar / 4\pi$. To numerically investigate the quasi-energy and quasi-eigenstate,
 we should use the finite truncation to approximate the $U_{m,n}$ matrix of infinite dimension~\cite{Longshi2017}. Such an approximated method is effective, since the matrix $U_{m,n}$ has the band structure~\cite{Izrailev87}.

\begin{figure}[htbp]
\begin{center}
\includegraphics[width=8.0cm]{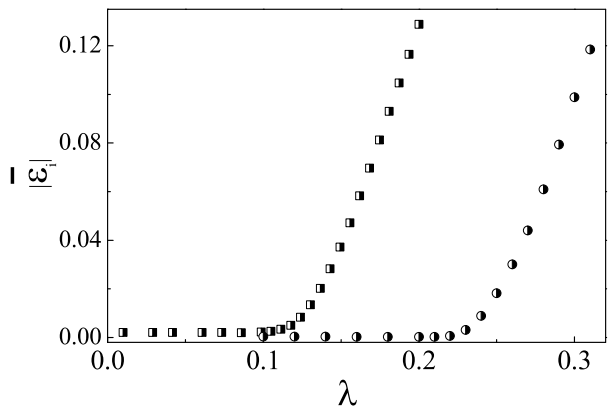}
\caption{(color online). The average value of the imaginary part of the quasi-energy, i.e., $|\bar{\varepsilon}_i|$ versus the strength of the complex potential $\lambda$ for $\ehbar=1.0$ (squares) and 1.5 (circles). The kick strength is $K=5$.\label{PTBKIrh}}
\end{center}
\end{figure}

The broken of $\mathcal{PT}$ symmetry is quantified by the appearance of the complex quasi-energy, i.e., $\varepsilon = \varepsilon_{\rm r} + i\varepsilon _{\rm i}$.
To identify such spontaneous symmetry breaking, we numerically calculate the average value of the imaginary part of the quasi-energy
    \begin{equation*}
    |\bar{\varepsilon}_{\rm i}|= \frac{1}{N}\sum_{j=1}^N |\varepsilon_{\rm i}^j| \;,
    \end{equation*}
where $N$ is the dimension of the Floquet matrix and $\varepsilon_{\rm i}^j$ denotes the imaginary part of the $j$-th the quasi-energy~\cite{Longshi2017}.
Our numerical results show that the average value $\bar{\varepsilon}_{\rm i}$ is virtually zero for small $\lambda$, and it abruptly increases once the $\lambda$ exceeds a certain critical threshold, i.e., $\lambda > \lambda_c$ (see Fig.~\ref{PTBKIrh}). This is a clear evidence of the spontaneous $\mathcal{PT}$-symmetry breaking controlled by the parameter $\lambda$. In numerical simulations, the truncation of the $U_{m,n}$ matrix  is $N = 2048$.
In fact, such $\mathcal{PT}$-symmetry breaking for the dynamical localization case has been reported in~\cite{Longshi2017}.

\begin{figure}[htbp]
\begin{center}
\includegraphics[width=8.0cm]{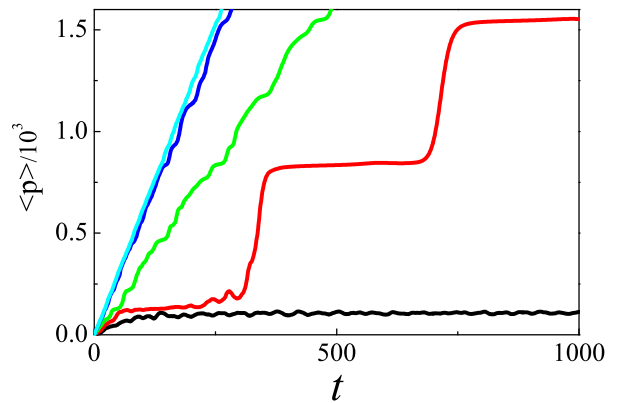}
\caption{(color online). Time dependence of the momentum current $\langle p \rangle$ with $\ehbar=1$ and $K=5.0$. From bottom to top $\lambda=0.06$ (black), 0.09 (red), 0.2 (green), 0.6 (blue) and 0.9 (cyan). \label{Current}}
\end{center}
\end{figure}

After studying the properties of eigenvalues of the Floquet operator, we are then to discuss the dynamical behaviour of the system, and we focus on the momentum current here. In the basis of $\vert \varphi_n \rangle$,  an arbitrary state can be expressed as
 $\psi(\theta,t)=\sum_{n=-\infty}^{+\infty}\psi_n(t) \langle\theta|\varphi_n\rangle$, with
 $\psi_n(t)$ being the wavefunction in the momentum representation.
The momentum current is quantified by the mean momentum
    \begin{equation*}
        \langle p(t) \rangle = \frac{\sum_n p_n |\psi_n(t)|^2}{\mtn}\;,
    \end{equation*}
where $\mtn = \sum_n |\psi_n(t)|^2$ is the norm of the quantum state~\cite{Longshi2017}.
For $\lambda > \lambda_c$, the appearance of the complex quasi-energies will lead to the  exponentially-fast increase of norm with time. The above definition of momentum current drops  the contribution from the growth of the norm to  the current behaviour.
We numerically investigate  the  momentum current for the irrational values of $\ehbar/4\pi$ with different $\lambda$.
In our numerical simulations, the initial state is taken as the ground state of angular momentum operator, i.e., $\psi(\theta,0)=1/\sqrt{2\pi}$.
Our numerical results show that, below the transition point, i.e., $\lambda< \lambda_c$ (see Fig.~\ref{Current} for  $\lambda=0.06$), the momentum current saturates to a small asymptotic value after the growth during the initially short time interval. In fact, in the limit of $\lambda \rightarrow 0$, the wavepackets spreads symmetrically in momentum space, thus the average momentum is virtually zero.
Interestingly, around the threshold value of spontaneous $\mathcal{PT}$-symmetry breaking, i.e., $\lambda \sim \lambda_c$ (e.g., $\lambda=0.09$ in Fig.~\ref{Current}), there is the staircase-growth of the momentum current. Moreover, the jump of the momentum current from the lower stair to the upper one is very sharp, which implies that the system changes to different quantum states with time evolution.
Above the transition point, i.e., $\lambda > \lambda_c$ (e.g., $\lambda=0.2$ in Fig.~\ref{Current}), the momentum current linearly increases with time, i.e., $\langle p(t) \rangle =Dt$, for which the growth rate $D$ is independent on $\lambda$ if $\lambda \gg \lambda_c$ (e.g., $\lambda=0.6$ and 0.9 in Fig.~\ref{Current}).
Our investigation on momentum current in dynamical localization regime may sight a new light on the understanding of the unidirectional  transport phenomenon.

\begin{figure}[htbp]
\begin{center}
\includegraphics[width=8.5cm]{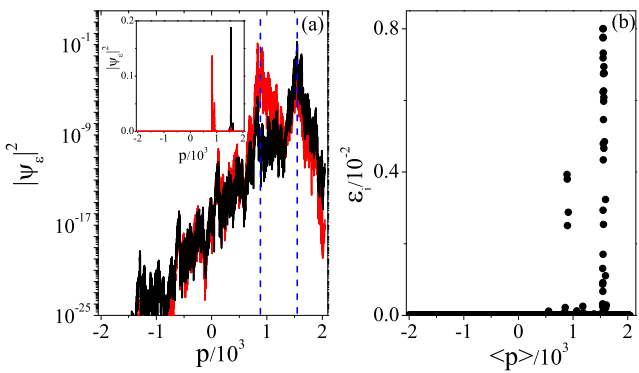}
\caption{(color online). (a) The Floquet eigenstates (in semi-log scale) in momentum space with $\varepsilon_{\rm i} = 0.00393$ (red) and  $0.008$ (black) which corresponds to the two peaks of $\varepsilon_{\rm i}$ in (b). The main plot and the inset show the same data in the lin-log and linear scales, respectively. Dashed lines (in blue) mark the center of each Floquet eigenstates.
(b) The imaginary part of the quasi-energy $\varepsilon_{\rm i}$ versus the mean momentum $\langle p_{\rm } \rangle$ of the corresponding eigenstate. The parameters are $K=5$, $\ehbar=1$, and $\lambda=0.09$.\label{Eigenstates}}
\end{center}
\end{figure}

The mechanism of the staircase growth of the momentum current can be understood as follows.
An arbitrary state can be expanded in the basis of the Floquet eigenstates. At the initial time, the expansion of the quantum state takes the form
        \begin{equation}\label{IniState}
        |\psi(t_0)\rangle =\sum_{\varepsilon} C_{\varepsilon}|\psi_{\varepsilon}\rangle\;,
        \end{equation}
where $|\psi_{\varepsilon}\rangle$ indicates the eigenstate of the Floquet operator and $C_{\varepsilon}$ is components of the quantum state.
After the $n$-th kicks, the quantum state has the expression
        \begin{equation}\label{AnyState}
        |\psi(t_n)\rangle =\sum_{\varepsilon} C_{\varepsilon}e^{-i n \varepsilon}|\psi_{\varepsilon}\rangle\;.
        \end{equation}
When the $\mathcal{PT}$ symmetry is broken, the quasienergy is complex
i.e., $\varepsilon = \varepsilon_{\rm r} + i \varepsilon_{\rm i}$. Therefore, the expansion of $|\psi(t_n)\rangle$ can be rewritten as
     \begin{align}
        |\psi(t_n)\rangle =
        \sum_{\varepsilon} C_{\varepsilon}e^{ n \varepsilon_{\rm i}}e^{-i n \varepsilon_{\rm r}}|\psi_{\varepsilon}\rangle
        \;.
     \end{align}\label{AnyState2}
It is apparent that the components with $\varepsilon_{\rm i}>0$ will exponentially grow.
Then, the mean momentum $\langle p_{} \rangle$ corresponding to these eigenstates contributes mainly to the momentum current. Interestingly, we find that most of the eigenstates with positive $\varepsilon_{\rm i}$ are localized in momentum space [see Fig.~\ref{Eigenstates}(a)]. Moreover, the mean momentum of those eigenstates concentrates on several separate values $p_m$ [see Fig.~\ref{Eigenstates}(b)]. Detailed observations show that each $p_m$ is in one-to-one correspondence  to the platform of the momentum current in Fig.~\ref{Current}. It is evident that, during the appearance of the platform of the momentum current, the quantum state is the eigentate of the largest $\varepsilon_{\rm i}$ with the same $\langle p \rangle$. The transition between the eigenstates of different mean momentum leads to the stair-case growth of the momentum current.

The underlying physics of the linear growth of the momentum current  for  $\lambda > \lambda_c$ is due  to the gain or loss mechanism induced by the imaginary part of the kicking potential. Such mechanism happens when the Floquet operator of the non-Hermitian term $U_{K}^{\rm i}(\theta)=\exp[K\lambda \sin(\theta)/\ehbar]$ operates on the quantum state, i.e.,  $U_{K}^{\rm i}\psi(\theta)$.
The action can dramatically cause the annihilation of the quantum state in the region of $\theta \in (-\pi,0)$, since in this region the value of $\sin(\theta)$ is negative. In contrast, the probability of the particle in the region of $(0,\pi)$ will be enhanced as $\sin(\theta)$ takes positive value in this region.
For $K\lambda/\ehbar \gg 1$, the value of the $U_K^{\rm i}(\theta)$ is extremely large in the position of $\theta=\pi/2$. Therefore,  the action of the Floquet operator $U_{K}^{\rm i}$ on a quantum state can effectively generate a quantum particle in $\theta=\pi/2$ if the center of the quantum state is not very far from this position. Indeed, our theoretical analysis proves that the wave function after each kicks can be well described by a Gaussion wave packet with the center $\theta_0 =\pi/2$~\cite{analysis}. In this position, the quantum particle experiences the kicking force of strength $K$. Therefore, the time growth of the mean momentum is roughly in the form of $\langle p(t) \rangle \propto Kt$.

\section{``Quantized'' acceleration mode}\label{sec3}

\begin{figure}[htbp]
\begin{center}
\includegraphics[width=7.5cm]{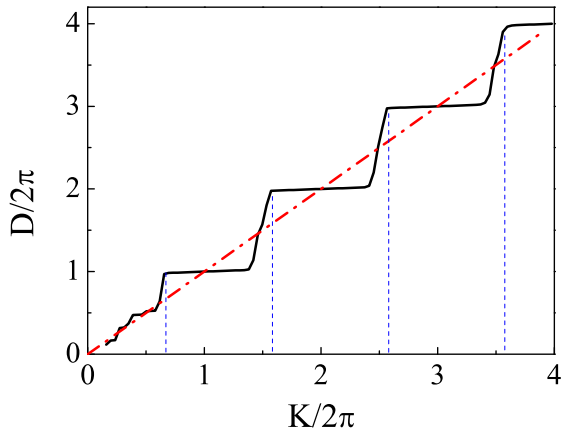}
\caption{(color online). The acceleration rate $D$ versus $K$ for $\ehbar=0.1$. Dashed-dotted line (in red) indicates the function of the form $D(K)=K$. Dashed lines (in blue) mark the transition points.\label{Dcoeff}}
\end{center}
\end{figure}

Further, we numerically investigate the acceleration rate of the momentum current, i.e., $D = \lim_{t\rightarrow t_f}\langle p^2(t)\rangle /t_f$ for $\lambda \gg \lambda_c$, where $t_f$ is the total time one can track the time evolution. Due to the linear growth of $\langle p(t)\rangle$, the $t_f$ of a scale of hundreds of
kicking periods can ensure the precise acceleration rate.
Interestingly, we find that the acceleration rate exhibits the ``quantized'' increment with increasing $K$ (see Fig.~\ref{Dcoeff}), namely, $D =  2n\pi $ for $K\in[2n\pi - \Delta_0, 2n\pi + \Delta_0]$ with $\Delta_0 \approx \pi$ and $n \ge 1$.
The mechanism of such an intriguing phenomenon is due to the coexistence of the classical acceleration mode and the ``gain-or-loss'' effects of the non-Hermitian potential. Remember that for $K\lambda/\ehbar \gg 1$ the action of the Floquet operator $U_{\lambda}$ on a quantum state can effectively generate a particle in the position of $\theta = \pi/2$. Indeed, our analytic analysis proves that, after the action of $U_{\lambda}$, the wavepackets  can be well described by a Gaussian function with minimum uncertainty $\delta\theta \delta p={\ehbar}/{2}$,  centered at $(\bar{\theta}= \pi/2,\bar{p}=2n\pi)$~\cite{analysis}. Then, we can regard it as a classical particle and analyse the acceleration mode of its classical trajectory.

We consider the classical acceleration mode of kicked rotor, which is governed by the classical mapping equation~\cite{Chirikov}
\begin{equation}\label{clmapping}
         \begin{cases}
            p(t_{j+1}) = p(t_j) + K \sin[\theta(t_j)]\;,\\
            \theta(t_{j+1})= \theta(t_j) + p(t_{j+1})\;,
         \end{cases}
\end{equation}
where $\theta(t_j)$ and $p(t_j)$ denote the angle coordinate and the angular momentum after the $j$-th kick. It is easy to see that a classical trajectory with $\left[\theta(t_0) = {\pi}/{2}, p(t_0)=2 m \pi\right ]$ will be accelerated linearly as time evolves, i.e., $\left[\theta(t_j) = {\pi}/{2}, p(t_j) = p_0 + jK\right ]$ if $K = 2 n \pi$, where $m$ and $n$ are all integers. In our model, we should also consider the effects of the imaginary part of the kicking potential on the time evolution of a classical trajectory.

Without loss of generality, we assume that at the time $t=t_j$ the position of a classical trajectory is $\left(\theta = {\pi}/{2}, p=0\right )$.
We consider the case that the kick strength has some deviation from the ideal value, i.e., $K= 2 n \pi + \Delta$ with $|\Delta| < \pi$. According to Eq.~\eqref{clmapping}, after one kick period, the trajectory changes to be
\begin{equation}\label{Ptj}
p(t_{j+1}^-) =2 n \pi + \Delta\;,
\end{equation}
and
\begin{equation}\label{Thetatj}
\theta(t_{j+1}^-)= \frac{\pi}{2} + 2n\pi + \Delta\;,
\end{equation}
where the superscript `-' indicates the time immediately before the action of imaginary part of the kicking potential, i.e., the $U_{K}^{\rm i}$ operator. Equation~\eqref{Thetatj} reveals that the value of $\Delta$ can be regarded as the distance between the center of the wavepackets and the position of $\theta = \pi/2 + 2n\pi$ which is essentially equal to $\pi/2$ due to the periodical boundary condition.
Since the action of  the $U_{K}^{\rm i}$ operator on the wavepackets greatly enhances the probability of a particle in $\theta = \pi/2  + 2n\pi$ if the value of $\Delta$  is smaller than a threshold value, i.e., $\Delta_0$, it is reasonable to believe that, after the action of $U_{K}^{\rm i}$, the particle moves  to the position of  $\theta(t_{j+1})= {\pi}/{2} + 2n\pi$.
Accordingly, its momentum becomes to be $p(t_{n+1}) =  2n\pi $ for which the actual increment of the momentum during one-period evolution is $D = 2n\pi$.
A rough estimation of $\Delta_0$ is a half of the width of the region $[2(n-1)\pi + \pi/2,2n\pi + \pi/2]$, i.e., $\Delta_0 \approx \pi$, which is confirmed by our numerical results in Fig~\ref{Dcoeff}.

\section{Summary}\label{sec4}

In this work, we investigate the directed current of the quantum kicked rotor model whose kicking potential satisfies the $\mathcal{PT}$ symmetric condition.
We find that in the vicinity of transition point, i.e., $\lambda \approx \lambda_c$, the eigenstates is well localized in momentum space. Moreover, the mean momentum eigenstates with positive real imaginary parts of the quasi-energy concentrates on several separate values. Such property leads to the staircase growth of the momentum current $\langle p(t) \rangle$ with time. When the parameter $\lambda$ is larger than the transition point, i.e., $\lambda > \lambda_c$, the momentum current linearly increases with time, i.e., $\langle p(t) \rangle = Dt$. We make extensive investigations on the acceleration rate $D$ for  $\lambda \gg \lambda_c$. Interestingly, we find that, for $K\lambda/\ehbar \gg 1$, the acceleration rate exhibits the ``quantized'' increment with the increase of $K$, i.e., $D =  2n\pi $ for $K\in [2n\pi - \Delta_0, 2n\pi + \Delta_0]$ with $\Delta_0 \approx \pi$ and $n \ge 1$. For $K\lambda/\ehbar \gg 1$, our analytic analysis proves that, at any time $t=t_n$, the wavepackets can be well described by the Gaussian funciton with a center $(\bar{\theta}=\pi/2, \bar{p}= 2n\pi)$. The motion of the wavepacket in phase space follows the classical acceleration mode of the trajectory of the kicked rotor model.
The theory of the modified acceleration mode of the classical particle by the gain-or-loss mechanism of the complex kicking potential can successfully  explain such ``quantized'' phenomenon of momentum current. It is known that cold atoms driven by time-periodical optical lattice, with rich and complex physics, such as the Butterfly
spectrum~\cite{Wang08}, the exponentially-fast diffusion~\cite{Wang11},  is an ideal platform for investigating the directed transport  phenomenon.
Our results may also be useful in the quantum control of the directed transport of matter waves.

\section*{ACKNOWLEDGMENTS}
This work was partially supported by the Natural Science Foundation of China under Grant Nos. 11447016, 11535011, 11775210, and 11575087.

\appendix

\section{The acceleration of momentum current}

The Floquet operator of the non-Hermitian QKR reads
\begin{equation}\label{FloqOPt}
U=U_{f}U_{K}^{\rm r}U_{K}^{\rm i}\;,
\end{equation}
with the free evolution operator  $U_f = \exp(-i p^2/2\ehbar)$, the evolution operator of the real part of the kicking potential
\begin{equation}\label{REPart}
\ U_{K}^{\rm r}=\exp\left[ -i\frac{K}{\ehbar}\cos (\theta)\right]\;,
\end{equation}
and that of the imaginary part
\begin{equation}\label{IMPart}
U_{K}^{\rm i}=\exp\left[\frac{\lambda K}{\ehbar}\sin (\theta)\right]\;.
\end{equation}
The maximum value of $U_{K}^{\rm i}$ corresponds to $\theta_0= \pi/2$.
In condition that  $\ehbar \rightarrow 0$ (with $\lambda K/\ehbar\gg1$ and $K/\ehbar\gg1$), the expansions of first order for  $U_{K}^{\rm i}$ and $U_{K}^{\rm r}$ around $\theta_0$ take the form
\begin{equation}\label{ExpanIMPart}
U_{K}^{\rm i}\approx \exp\left[-\frac{\lambda K(\theta-\theta_0)^{2}}{2\ehbar}\right]\exp\left(\frac{\lambda K}{\ehbar}\right)\;,
\end{equation}
and
\begin{equation}\label{ExpanREPart}
U_{K}^{\rm r}\approx \exp\left[\frac{iK(\theta-\theta_0)}{\ehbar}\right]\;.
\end{equation}
As a further step, we consider the time evolution of a quantum state under the action of above operators.

Without loss of generality, we assume that the initial state is a Gaussian wavepacket
\begin{equation}\label{InitSta}
\psi(\theta,t_{0})=\frac{1}{(\sigma^2 \pi)^4}\exp\left(-\frac{\theta^{2}}{2\sigma^{2}}+\frac{ip_{0}\theta}{\ehbar}\right)\;,
\end{equation}
for which the uncertainty relation is such that $\delta \theta \delta p = \ehbar/2$.
The evolution of quantum states from $t_0$ to $t_{1} = t_0+1 $ is governed by $|\psi (t_{1})\rangle = U |\psi (t_{0})\rangle $. The action of $U_{K}^{\rm i}$ on $\psi(\theta,t_{0})$ yields that
\begin{equation}\label{InitSta2}
\widetilde{\psi_{}}(\theta,t_{0}) \propto
\exp\left[-\frac{\lambda K(\theta-\theta_0)^{2}}{2\ehbar}-\frac{\theta^{2}}{2\sigma^{2}}+\frac{ip_{0}\theta }{\ehbar}\right]\;.
\end{equation}
In condition that  $\lambda K/\ehbar\gg 1/\sigma^2$, we can neglect the contribution of the second term in right-hand side of the above equation. Then, the quantum state can be approximated as
\begin{equation}\label{InitSta3}
\widetilde{\psi_{}}(\theta,t_{0}) \propto
\exp\left[-\frac{\lambda K(\theta-\theta_0)^{2}}{2\ehbar}+\frac{ip_{0}\theta }{\ehbar}\right]\;.
\end{equation}
For the convenience of analysis, hereafter, we use this state as the initial state for the time evolution. That is to say the Floquet operator is redefined as
\begin{equation}\label{FloqOPt2}
U=U_{K}^{\rm i}U_{f}U_{K}^{\rm r}\;.
\end{equation}
By comparison Eq.~\eqref{FloqOPt2} with Eq.~\eqref{FloqOPt}, one can see that the sequence of the action of $U_{K}^{\rm i}$ and $U_{f}$ in time evolution is rearranged, which actually has no effect on physical results.

Consider the action of $U_{K}^{\rm r}$ in Eq.~\eqref{ExpanREPart} on
$\widetilde{\psi}(\theta,t_{0})$,
\begin{align}\label{StateK+}
\psi(\theta,t_{0}^+) &=U_{K}^{\rm r}\widetilde{\psi_{}}(\theta,t_{0})\\\nonumber
&\propto\exp\left[-\frac{\lambda K}{2\ehbar}(\theta-\theta_0)^{2}+\frac{iK}{\ehbar}(\theta-\theta_0)\right]\\\nonumber
& \hspace{30pt}\times \exp\left(\frac{i}{\ehbar}p_{0}\theta\right)\;,
\end{align}
where the superscript `+' indicates the time immediately after the action of the real part of the kicking potential. Next step is the action of the free evolution operator $U_{f}$ on the quantum state. Before that we should transform the state to momentum space,
\begin{align}\label{}
\psi(p,t_{0}^+) &=\int_{-\pi}^{\pi}\psi_{}(\theta,t_{0}^+)\exp(-ip\theta/\ehbar)d\theta\\\nonumber
& \propto\exp\left[-\frac{(p-p_K)^{2}}{2\ehbar\lambda K}-\frac{i p \theta_0}{\ehbar}\right]\;,
\end{align}
where $p_K = p_0+K$. Then, the action of $U_{f}$ on $\psi(p,t_{0}^+)$ yields
\begin{align}\label{UPActionState}
\psi(p,t_{1}^{-})& = U_{f}\psi(p,t_{0}^+)\\\nonumber
&\propto\exp\left[-\frac{(p-p_{K})^{2}}{2\ehbar\lambda K}-\frac{ip(p+2\theta_0)^{}}{2\ehbar}\right]\;,
\end{align}
where the superscript `-' indicates the time immediately before the action of the kicking operator $U_{K}^{\rm i}$.

The corresponding quantum state in real space is
\begin{align}\label{FFTX}
\psi_{}(\theta,t_{1}^-) & \propto\int_{-\infty}^{\infty}dp\psi_{}(p,t_{1}^-)
\exp(ip\theta/\ehbar)\\\nonumber
& \propto \exp\left[-\frac{\left(\theta-\theta_0-p_{K} \right)^{2}}{2\ehbar\lambda K}\right]\\\nonumber
&\hspace{30pt} \times \exp\left[\frac{i\left(\theta-\theta_0+\frac{p_{K}}{\lambda^{2}K^{2}}\right)^{2}}{2\ehbar}\right]\;.
\end{align}
We assume that
\begin{equation}\label{Kcondition}
p_K=2n_{0}\pi+\Delta\;,
\end{equation}
with $-\pi < \Delta < \pi$.
Then, the quantum state in Eq.~\eqref{FFTX} is rewritten  as
\begin{align}\label{FFTX2}
\psi_{}(\theta,t_{1}^-) \propto
& \exp\left\{-\frac{\left[\theta-(\theta_0+2n_0 \pi)-\Delta \right]^{2}}{2\ehbar\lambda K}\right\}\\\nonumber
&\hspace{20pt} \times \exp\left[\frac{i\left(\theta-\theta_0+\frac{p_{K}}{\lambda^{2}K^{2}}\right)^{2}}{2\ehbar}\right]\;.
\end{align}
Apparently, Eq.~\eqref{FFTX2} indicates a Gaussian wavepackets with the center $\bar{\theta} = \theta_0 + 2n\pi + \Delta$.
If the distance between $\bar{\theta}$ and $\theta_0 + 2n\pi$ is smaller than a threshold value $\Delta_0$, the action of ${U}_{K}^{\rm i}$ (in Eq.~\eqref{ExpanIMPart}) on the quantum state $\psi_{}(\theta,t_{1}^-)$ can
effectively enhance the probability of a particle in the position of $\bar{\theta} = \theta_0 + 2n\pi$. Then, we get
\begin{align}\label{1TState}
\psi_{}(\theta,t_{1}^{+}) = & {U}_{K}^{\rm i}
\psi_{}(\theta,t_{1}^-)\\\nonumber
\propto
& \exp\left\{-\frac{\left[\theta-(\theta_0+2n_0 \pi) \right]^{2}}{2\ehbar\lambda K}\right\}\\\nonumber
&\hspace{20pt} \times \exp\left[\frac{i\left(\theta-\theta_0+\frac{p_{K}}{\lambda^{2}K^{2}}\right)^{2}}{2\ehbar}\right]\;,
\end{align}
where the superscript `+' indicates the time immediately after the action of ${U}_{K}^{\rm i}$.
A rough estimation of $\Delta_0$ is a half of the width of the region $[2(n-1)\pi + \theta_0,2n\pi + \theta_0]$, i.e., $\Delta_0 \approx \pi$.
Now, the time evolution of a quantum state during  an entire kicking period ends.

It is apparent that the center of this wavepacket in real space is \begin{equation}\label{MeanX}
\overline{\theta}_{t_1}=2n_{0}\pi+ \theta_0\;.
\end{equation}
with the second moment
\begin{equation}\label{XSecondM}
\delta \theta=\sqrt{\overline{\theta^{2}}-(\overline{\theta})^{2}}=\sqrt{\frac{\ehbar}{2\lambda K}}\;.
\end{equation}
To obtain the momentum center of the wavepacket $\psi_{}(\theta,t_{1}^{+})$, we should transform it to momentum space,
\begin{align}\label{GaussianP4}
\psi_{}(p,t_{1}^+)\propto&
\int d\theta e^{-ip \theta/\hbar}\psi_{}(\theta,t_{1}^+)\\\nonumber
\propto & \exp\left[\frac{(p-2n_{0}\pi)^{2}}{2\hbar\lambda K}+if(p)\right]\;,
\end{align}
where $f(p)$ is an unimportant function of momentum. The function $f(p)$ does not determine the mean momentum, hence it is no need to know the specific form of $f(p)$. From the above quantum state, we can get the mean momentum
\begin{equation}\label{MeanP}
\overline{p}_{t_1}=2n_{0}\pi\;.
\end{equation}
with the second moment
\begin{equation}\label{PSecondM}
\delta p=\sqrt{\overline{p^{2}}-(\overline{p})^{2}}=\sqrt{\frac{\ehbar\lambda K}{2}}\;.
\end{equation}
One can find that the quantum state $|\psi (t_1)\rangle$ satisfies the uncertainty relation
\begin{equation}\label{Uncerta}
\delta \theta \delta p=\frac{\ehbar}{2}\;,
\end{equation}
which is same as that of the initial wavepacket.

As a brief summary, the quantum state after the evolution of first kicking period can be well described by a Gaussian wavepacket,
\begin{equation}\label{FTState}
\psi_{}(\theta,t_{1})\simeq \exp\left[-\frac{\lambda K(\theta-\bar{\theta}_{t_{1}})^{2}}{2\ehbar}+\frac{i\bar{p}_{t_{1}}\theta}{\ehbar}\right]\;,
\end{equation}
where
\[
\begin{cases}
\bar{\theta}_{t_{1}}=2n_{0}\pi+\theta_0\;,\\
\bar{p}_{t_{1}}=2n_{0}\pi\;,
\end{cases}
\]
with $\theta_0 = \pi/2$.
The time evolution of quantum state for $t> t_1$ just repeats the above procedure. Accordingly, the quantum state at any time $t=t_n$ can be well approximated by a Gaussian wavepacket. Moreover, we can get the its center $(\bar{\theta}_{t_n},\bar{p}_{t_n})$ by using the iterative method.

Firstly, we can get the center $(\bar{\theta}_{t_2},\bar{p}_{t_2})$ at the time $t=t_2$.
Taking accounting to $\bar{p}_{t_1}= 2 n_0 \pi$ [see Eq.~\eqref{MeanP}], for $K = 2 n \pi + \Delta$ ($-\pi < \Delta< \pi$),  then we arrive at $\bar{p}_{t_1}+ K = 2 (n_0 + n) \pi+ \Delta$. Note that the value of $\Delta$ is smaller than a threshold value $\Delta_0$.
By repeating the procedure of the derivation for $(\bar{\theta}_{t_1},\bar{p}_{t_1})$, one obtains that
\begin{equation}\label{Center2}
\begin{cases}
\bar{\theta}_{t_2} = 2(n_0 + n )\pi + \theta_0\;,\\
\bar{p}_{t_2} = 2(n_0 + n  )\pi\;.
\end{cases}
\end{equation}
By using the same method, one can find
\begin{equation}\label{CenterN}
\begin{cases}
\bar{\theta}_{t_{j}}= 2\left[n_{0} + (j-1)n\right]\pi+\theta_0\;,\\
\bar{p}_{t_{j}}= 2\left[n_{0} + (j-1)n\right]\pi\;.
\end{cases}
\end{equation}
It is evident that the acceleration rate is
\begin{equation}\label{GRate}
D = \bar{p}_{t_{j}}-\bar{p}_{t_{j-1}}=2n_{}\pi\;,
\end{equation}
for $K \in [2n\pi - \Delta_0, 2n\pi + \Delta_0]$ with $\Delta_0\approx \pi$. Our analytic analysis is confirmed by numerical results.

\end{document}